\documentstyle[12pt,epsf,aasms4]{article}
\newcommand{\etal}{et~al.}

\lefthead{Kundi\'c et~al.}
\righthead{0957+561 Time Delay and $H_0$}

\begin{document}

\title{A Robust Determination of the Time Delay in 0957+561A,B and a
Measurement of the Global Value of Hubble's Constant}

\author{Tomislav Kundi\'c}
\affil{Theoretical Astrophysics, California Institute of Technology,
Mail Code 130-33, Pasadena, CA 91125}
\and
\author{Edwin L. Turner, Wesley N. Colley\altaffilmark{1}
\altaffiltext{1}{Supported by the Fannie and John Hertz Foundation},
J. Richard Gott, III, James E. Rhoads\altaffilmark{2},
\altaffiltext{2}{Currently at the Kitt Peak National Observatory} Yun
Wang} 
\affil{Princeton University Observatory, Peyton Hall, Princeton, NJ
08544} 
\affil{lens@astro.princeton.edu}
\and
\author{Louis E. Bergeron\altaffilmark{3}\altaffiltext{3}{Currently at
the Space Telescope Science Institute}, Karen A. Gloria, Daniel
C. Long}
\affil{Apache Point Observatory, 2001 Apache Point Rd., P. O. Box 59,
Sunspot, NM 88349}
\and
\author{Sangeeta Malhotra}
\affil{Infrared Processing and Analysis Center, California Institute
of Technology, Mail Code 100-22, Pasadena, CA 91125} 
\and
\author{Joachim Wambsganss}
\affil{Astrophysikalisches Institut Potsdam, An der Sternwarte 16,
14482 Potsdam, Germany}

\begin{abstract}

Continued photometric monitoring of the gravitational lens system
0957+561A,B in the $g$ and $r$ bands with the Apache Point Observatory
(APO) 3.5 m telescope during 1996 shows a sharp $g$ band event in the
trailing (B) image light curve at the precise time predicted in an
earlier paper. The prediction was based on the observation of the
event during 1995 in the leading (A) image and on a differential time
delay of 415 days.  This success confirms the so called ``short
delay'', and the absence of any such feature at a delay near 540 days
rejects the ``long delay'' for this system, thus resolving a long
standing controversy.  A series of statistical analyses of our light
curve data yield a best fit delay of $417 \pm 3$ days (95\% confidence
interval) and demonstrate that this result is quite robust against
variations in the analysis technique, data subsamples and assumed
parametric relationship of the two light curves.

Recent improvements in the modeling of the lens system (consisting of
a galaxy plus a galaxy cluster) allow us to derive a value of the
global (at $z = 0.36$) value of Hubble's constant $H_0$ using
Refsdal's method, a simple and direct (single step) distance
determination based on experimentally verified and securely understood
physics and geometry.  The result is $H_0 = 64 \pm 13$ km/s/Mpc (for
$\Omega = 1$) where this 95\% confidence interval is dominantly
due to remaining lens model uncertainties.  However, it is reassuring that
available observations of the lensing mass distribution over constrain
the model and thus provide an internal consistency check on its
validity.  {\em We argue that this determination of the extragalactic
distance scale (10\% accurate at 1$\sigma$) is now of comparable quality, 
in terms of both
statistical and systematic uncertainties, to those based on more
conventional techniques.}

Finally, we briefly discuss the prospects for improved $H_0$ determinations
using gravitational lenses and some other possible implications and uses
of the 0957+561A,B light curves.

\end{abstract}


\section{Introduction \label{intro.sec}}

More than 30 years ago, Refsdal (1964, 1966) pointed out that the
differential light propagation time delay between two or more
gravitationally lensed images of a background object establishes an
absolute physical distance scale ($c \Delta t$) in the system. Thus,
the distance to a high-redshift object is directly measured, yielding
a value of Hubble's constant, $H_0$. The theory of this technique has
been elaborately developed and its realization has become a major
focus of gravitational lens studies [Narayan (1991) gives an
especially elegant treatment; see Blandford \& Narayan (1992) and
Narayan \& Bartelmann (1996) for reviews].

It may be useful to briefly review the main strengths of 
the lensing method in determination of the extragalactic distance
scale: 
\begin{enumerate}
\item It is a geometrical method based on the well understood and
experimentally verified physics of General Relativity in the
weak-field limit. By contrast, most conventional astronomical
techniques for measuring extragalactic distances rely either on 
empirical relationships or on our understanding of
complex astrophysical processes, or both.
\item It provides a direct, single step measurement of $H_0$ for each
system and thus avoids the propagation of errors along the ``distance
ladder'' which is no more secure than its weakest rung.
\item It measures distances to cosmologically distant objects, thus
precluding the possibility of confusing a local with a global
expansion rate.  Note that both observed CMB fluctuations and COBE
normalized numerical simulations of large-scale structure formation
suggest the possibility of 10-20\% {\em rms} expansion rate
fluctuations even on scales of order 10,000 km/s for some cosmological
models (D. Spergel 1996, private communication; Turner, Cen \&
Ostriker 1992).
\item Independent measurement of $H_0$ in two or more lens systems
with different source and lens redshifts allows a powerful internal
consistency check on the answer.  Although an inaccurate model of the
lens mass distribution or other systematic problems could yield an
incorrect distance in any particular system, no known or imagined
problem will consistently give {\em the same} wrong answer when
applied to different lenses.  Thus, if a small number of time delay
measurements all give the same $H_0$, this value can be regarded as
correct with considerable confidence.
\end{enumerate}

Despite these potent virtues, the practical realization ({\em i.e.},
at a useful and competitive accuracy) of Refsdal's method for
measuring $H_0$ has proven quite challenging and has been long
delayed.  For the lens system 0957+561A,B (Walsh, Carswell \& Weymann
1979), by far the best studied case, there are two basic reasons.
First, there has been sufficient ambiguity in detailed models of the
mass distribution in the lensing galaxy and associated cluster to
allow values of $H_0$ different by a factor of two or more to be
consistent with the same measured time delay (Young \etal\ 1981;
Narasimha, Subramanian \& Chitre 1984; Falco, Gorenstein \& Shapiro
1991; Kochanek 1991); fortunately, this problem has been much
alleviated by recent theoretical and observational work.  See
\S~\ref{model.sec} for details.  Second, despite extensive optical
(Lloyd 1981; Keel 1982; Florentin-Nielsen 1984; Schild \& Cholfin
1986; Vanderriest \etal\ 1989; Schild \& Thomson 1995) and radio
(Leh\'ar \etal\ 1992; Haarsma \etal\ 1996) monitoring programs
extending over a period of more than 15 years, values of the
differential time delay discrepant by more than 30\% have continued to
be debated in the literature.  In particular, most studies have given
a delay either in the range 400--420 days or one of about 530--540
days.  These two rough values, the ``short delay'' and the ``long
delay'' have been obtained {\em both by applying the same statistical
techniques to different data sets and by applying different
statistical techniques to the same data} [Vanderriest \etal\ 1989;
Leh\'ar \etal\ 1992; Press, Rybicki \& Hewitt 1992a, 1992b (hereafter
collectively referred to as PRH); Pelt \etal\ 1994, 1996 (hereafter
collectively PHKRS); Beskin \& Oknyanskij 1995]!  Moreover, even
application of a single technique to a single (radio) data set has
produced best estimate delays that move far outside the nominal formal
high confidence interval as additional points in the light curve
accumulate (Press \etal\ 1992b, Haarsma \etal\ 1996). The history of
the 0957+561A,B time delay, which can certainly be described as
confusing and controversial, is reviewed by Haarsma \etal\ (1996).

In this paper we report a robust determination of the time
delay which we believe effectively resolves the controversy in favor
of the short delay.  In addition, we use this delay and the results of
recent theoretical (Grogin \& Narayan 1996, hereafter GN) and
observational (Garrett \etal\ 1994) studies of the lens mass
distribution to calculate a global measure of Hubble's constant of
accuracy comparable to that of the best conventional techniques, both
in terms of statistical and systematic errors.

Our time delay determination differs from all previous ones in that
the appearance of a sharp, large amplitude feature in the $g$ band
light curve of the trailing image (0957+561B) during 1996 was {\em
predicted in advance} based on observations of the leading image
(0957+561A).  The light curve data showing this sharp $g$ band
event, plus other weaker features in the $g$ and $r$ band light
curves, is given in Kundi\'c \etal\ (1995, hereafter Paper I) along
with predictions of when it would appear in the B image during 1996
for either the short or long delay.  This paper reports 1996 data
showing that the short time delay prediction was quantitatively
correct while the long time delay prediction is not even qualitatively
consistent with the 1996 image B light curve.  Oscoz \etal\ (1996)
have very recently employed the Paper I predictions to exclude a delay
longer than 500 days, a result entirely consistent with those presented
here.

\S~\ref{data.sec} presents the 1996 light curve data for both images in
which the short delay predictions are clearly confirmed.
\S~\ref{stat.sec} presents a set of statistical analyses of the light
curves from which we derive a best fit value of the delay and estimate
its uncertainty.  In \S~\ref{model.sec}, we derive a global value of
Hubble's constant $H_0$ from the measured delay and discuss its
statistical and systematic uncertainties.  Finally, in
\S~\ref{summary.sec}, we comment on some other implications of the data
and on the general situation in attempts to apply Refsdal's method for
determining $H_0$.

\section{Photometric Data and Confirmed Prediction \label{data.sec}}

The 0957+561 photometric monitoring program at the Apache Point
Observatory (APO\footnote{APO is privately owned and operated by the
Astrophysical Research Consortium (ARC), consisting of the University
of Chicago, Institute for Advanced Study, Johns Hopkins University, New
Mexico State University, Princeton University, University of
Washington, and Washington State University.})  3.5 meter telescope
which is described in Paper I for the late 1994 to mid-1995 season
(hereafter referred to as the 1995 season) was continued in the late
1995 to mid-1996 season (hereafter designated as the 1996 season).
The instrumentation, filter system, observing protocols, and data
reduction techniques were as described in Paper I except that the large
majority of the 1996 observations were made from the APO site rather
than remotely from Princeton.  The 1996 light curves consist of 51 $g$
band and 54 $r$ band brightness measurements made from 25 November 1995
to 6 July 1996, as compared to the 46 $g$ and 46 $r$ band data points
obtained from 2 December 1994 to 27 May 1995 in the 1995 season. In
1996, the mean photometric error of a nightly measurement was 10
millimag in $g$ band and 9 millimag in $r$ band, compared to respective
values of 12 and 12 in the previous season.  The light curves for the
two seasons are thus of quite similar quality.

\begin{figure}
\epsscale{0.8}
\plotone{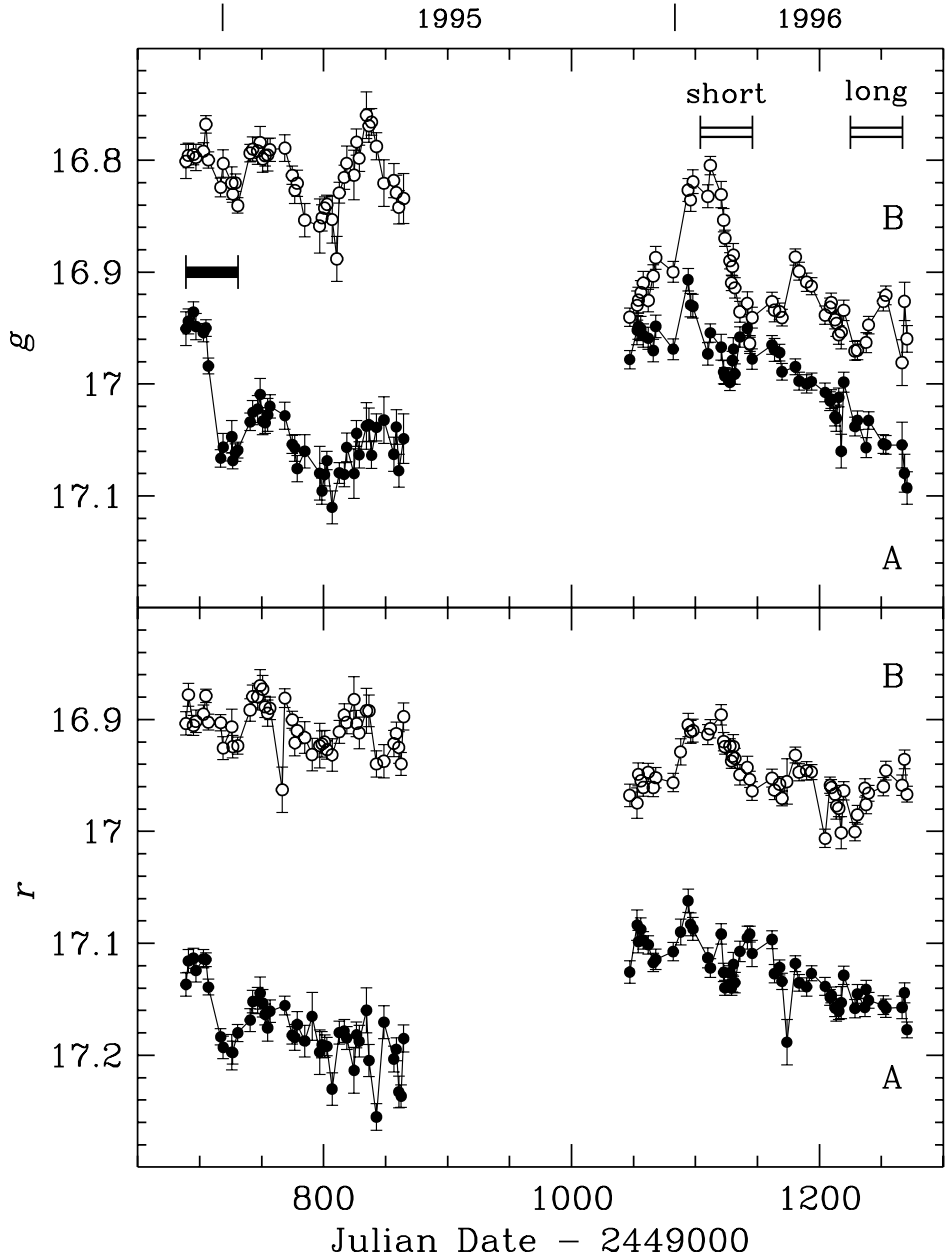}
\caption{The 1995 and 1996 $g$ and $r$ band light curves (with
1$\sigma$ error bars) for images A (filled points) and B (open points)
of the gravitationally lensed quasar 0957+561. The sharp event
observed at the beginning of the 1995 season (December 94--January 95)
in the $g$ light curve of the leading (A) image is marked with a
horizontal bar. Based on previously reported time delays in this
system, two predictions for the 1996 B light curve were presented in
Paper I; they are also marked (February and June 96).  The data
provide compelling evidence in favor of the shorter delay, thus
resolving a long-standing controversy.}
\label{alldata.fig}
\end{figure}

Figure~\ref{alldata.fig} displays our $g$ and $r$ data for images A and
B in the 1995 and 1996 seasons, in other words our complete photometric
data set.  This data is available in the {\tt elt/:0957} subdirectory
of the anonymous ftp area at {\tt astro.princeton.edu}.

The predictions of the short time delay set forth in Paper I can be
confirmed by inspection of Fig.~\ref{alldata.fig} without need for the
detailed statistical analysis presented in the next section.  The sharp
event which Paper I noted in the $g$ band light curve of image A in
early 1995 is marked in Fig.~\ref{alldata.fig}, as are the times of its
predicted appearance in the 1996 image B light curve.  An event of just
the expected amplitude and duration is seen in image B at the time
corresponding to the short delay, and no such feature is seen at a time
corresponding to the long delay.  Although we will proceed to a
quantitative analysis below, the qualitative success of the short delay
prediction is compelling.

Note that Fig.~\ref{alldata.fig} also displays the rather quiescent
behavior of image A in 1996, which corresponds to the predicted 1997
behavior of image B, modulo the now determined delay and magnitude
offset.  It also emphasizes the good fortune of the occurrence of the
sharp $g$ band event in image A shortly after the beginning of
our monitoring program; no comparable feature has been seen since!

Also note that photometric events in the $r$ band are consistently
smaller in amplitude than those in the $g$ band.  Since most previous
monitoring programs have concentrated their attention in red bands
(probably to minimize problems with moonlight), this may explain some
of the difficulty in obtaining a definitive time delay measurement.

\section{The Best Fit Delay and Its Uncertainty \label{stat.sec}}

In this section, we present a series of statistical comparisons of our
1995 image A light curve with the 1996 image B light curve.  The goals
of these analyses are 1) to determine the best estimate differential
time delay for 0957+561A,B; 2) to quantitatively assess its error (95\%
confidence interval); and 3) to demonstrate that the data provide a
{\em robust} delay measurement.  This last point deserves brief
elaboration: Previous attempts to measure the 0957+561A,B time delay
from radio and optical data have often depended as much on the
statistical techniques employed as upon the data, {\em i.e.}, different
methods gave different answers when applied to the same data (see
references in \S~\ref{intro.sec}).  Moreover, even the best delay as
determined by a single method was unstable against the inclusion or
exclusion of a few data points in some cases (PHKRS).  We demonstrate
below that the data shown in Fig.~\ref{alldata.fig} produce no such
difficulties or ambiguities.

\begin{figure}
\plotone{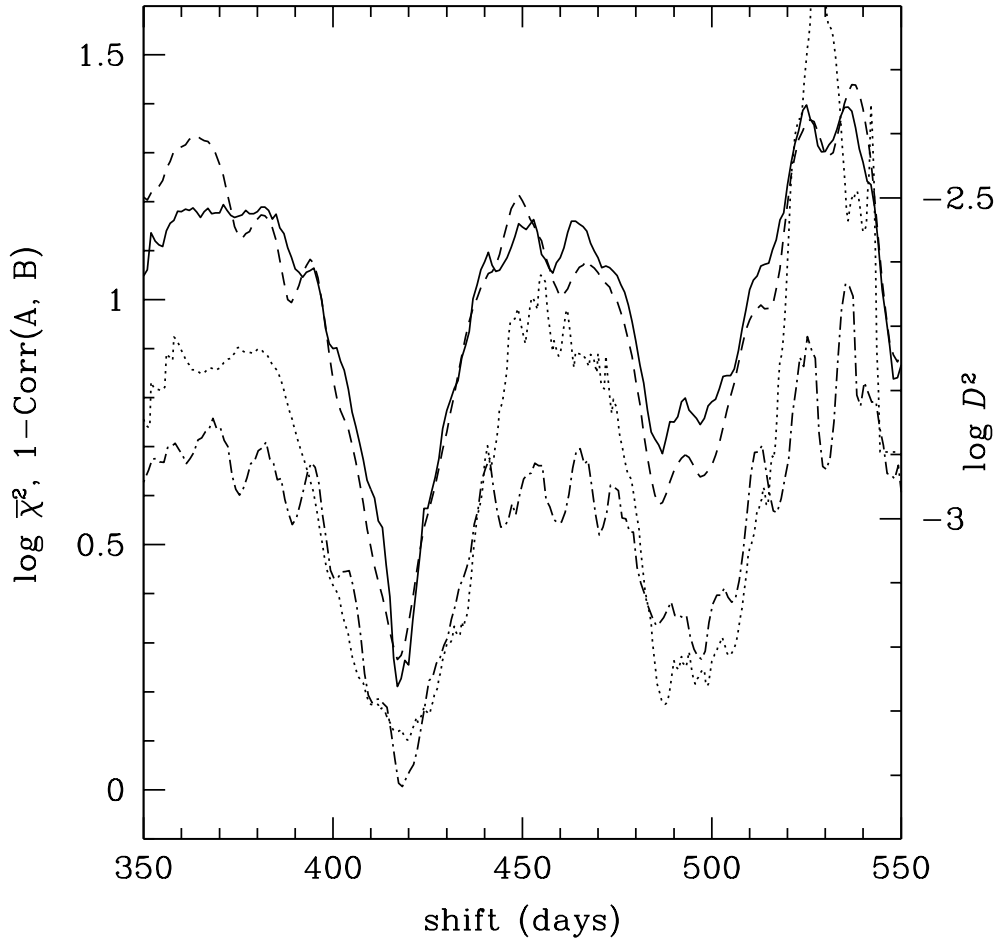}
\caption{Figure of merit for various statistical methods as a function
of time delay $\Delta t$ based on $g$ band light curves.  Curves are
arranged so that their minima correspond to the best fit delays. The
second fitted parameter, magnitude offset $\Delta m$, was minimized at
each value of $\Delta t$. The values of $\bar{\chi}^2$ for the
linear and PRH methods are represented with solid and dot-dash lines
respectively; the dispersion measure $D^2$ of the PHKRS method is
dashed, and the complement of the cross-correlation coefficient is
dotted.  Note that three different vertical coordinates are
represented, depending on the statistical technique.  All methods give
minima at $\sim$ 417--420 days and strongly reject a delay of $\sim$
540 days.}
\label{chi2.fig}
\end{figure}

Table~1 and Fig.~\ref{chi2.fig} summarize the results of our
statistical comparisons of the 1995 image A light curve with the 1996
image B data.  We employ four different statistical methods on four
subsets of the complete data set and four parameterized models of how
the image A and B light curves are related, although we do not present
all 64 of the possible combinations here.

Of the at least 11 different statistical analyses previously applied
to 0957+561A,B optical and radio light curves by various authors
(tabulated by Haarsma \etal\ 1996), we have selected four
representative ones intended to span the range of reasonable
approaches for analysis of our data.
\begin{enumerate}
\item Linear method: First and simplest, each light curve and its
errors are linearly interpolated, and the data points from the other
light curve are shifted so as to minimize $\bar{\chi}^2$ ($\chi^2$ per
degree of freedom). The number of degrees of freedom is equal to the
number of overlapping data points in the shifted light curves minus the
number of fitted parameters. Time delays with a small overlap of data
are thus penalized with respect to delays where the overlap is
significant. The 95\% confidence interval quoted in Table~1 for each
fitted parameter is derived by bootstrap resampling of residuals from
the combined light curve smoothed by a 7-point triangular filter with
a maximum length of 28 days. The confidence interval is robust with
respect to the choice of the smoothing filter.
\item PRH method: As emphasized by Press \etal\ (1992a, b), 
interpolation of the light curves is a crucial ingredient in any
method, and we adopt their rather elegant, though somewhat assumption
laden, interpolation scheme plus $\chi^2$ minimization as our second
method. The PRH method relies heavily on a model of the structure
function of the source's intrinsic variability.  We have taken a
(measured) structure function of $\log[C_{g}(\tau)] = -4.3 +
0.86\tau$; $\log[C_{r}(\tau)] = -4.5 + 0.83\tau$, where $\tau$ is in
days and $C$ is in magnitudes squared.  The quoted error on the PRH
scaled $\chi^2$ method is the $\Delta \chi^2$ interval of $\pm 4$.
While the association of a 95\% (or any other specific) confidence
interval with this $\Delta \chi^2$ interval is problematic, it
provides a rough estimate of the measurement error; we refer readers
to PRH for further discussion.
\item PHKRS method: As an alternative to $\chi^2$ minimization and as
a more assumption-free approach, we adopt a non-parametric technique
suggested by PHKRS as a third method. In
particular, we use their dispersion measure $D_{4, k}^2$ from the
second paper with linearly decreasing weights $S_{n, m}^{(2)}$ and
decorrelation length of $\delta = 7$ days. Here again 95\% confidence
intervals quoted in Table~1 are derived from the bootstrap resampling
of residuals.
\item Cross-correlation method: Finally, we have employed a
conventional cross-correlation analysis, similar to that used by
Schild \& Cholfin (1986), Vanderriest \etal\ (1989), Schild (1990),
and Leh\'ar \etal\ (1992), as our fourth method; this
technique also requires interpolation but seems to be rather
insensitive to its details.  Here we report cross-correlation results
using only the simplest linear interpolation scheme.  The errors in
Table 1 for this method are estimated 1$\sigma$ bounds on the
position of the correlation peak based on a bootstrapping analysis.
\end{enumerate}

The statistical methods described above have been applied to four
subsets of our light curve data.  First, we have considered only the
$g$ band data; we believe that this gives the best and most reliable
delay due to the high signal-to-noise provided by the sharp event.
Second, we have treated the $r$ band data separately; it provides a
partially independent, though lower signal-to-noise, delay
determination.  Third, we have combined the $g$ and $r$ data to find
overall best fit parameter values.  Fourth, we have used only the JD
2449689--2449731 interval of the 1995 image A light curve, {\em i.e.},
the sharp $g$ band event alone, to search the 1996 image B light curve
for a best fit.  This last subsampling of the data corresponds most
closely to testing the Paper I prediction and is not an {\em a
posteriori} (an hence invalid) editing of the data only because of
that prediction.

The simplest model of the mapping of the A image light curve into that
of the B image is just a fixed offset in time, $\Delta t$, and a fixed
magnitude offset, $\Delta m$. These two parameters correspond to the
macrolensing differential delays and image magnifications.  We also
consider models in which the magnitude offset is allowed to be a
linear function of time at a rate $\mu$, motivated by the possibility
that one or both images are being affected by microlensing events with
characteristic time scales long compared to the delay and the extent
of the observed light curves (Chang \& Refsdal 1979; Young 1981; Gott
1981; Falco, Wambsganss \& Schneider 1991).  Finally, we know that the
image B photometry is contaminated by the light of the lensing galaxy
G1, especially in the $r$ band due to the galaxy's red color relative
to the quasar source. Thus, we consider a parameter $m_{\rm gal}$
which represents a constant flux added to that of the source in image
B.  The general form of our model of the relation of the image A light
curve to that of image B is, therefore
\begin{equation}
10^{-0.4 m_B(t)} = 10^{-0.4 \left[ m_A(t - \Delta t) + \Delta m + \mu (t -
t_0) \right]} + 10^{-0.4 m_{\rm gal}} \quad ,
\end{equation}
where $m_A$, $m_B$ are the measured image A and B magnitudes; $\Delta
t$ is the time delay; $\Delta m$, $\mu$ are the magnitude offset and
its rate of change; $t_0$ = JD 2450000 (arbitrarily chosen); and
$m_{\rm gal}$ is the magnitude of the lensing galaxy G1 which
contaminates image B photometry. As shown in Table~1, we have
experimented with models in which up to three parameters are allowed
to vary as well as ones in which the less interesting ones, $\mu$ and
$m_{\rm gal}$ are neglected, that is fixed at zero and infinity,
respectively.

Table~1 shows the results of a selected set of combinations of our
statistical methods, light curve mapping models and versions of the
data.  All 64 possible combinations are not shown for the sake of
clarity and because some of the statistical methods and some versions
of the data do not lend themselves well to the determination of some
parameters.  However, we emphasize that we are aware of no combination
of method, data subset and model which gives a $\Delta t$ value significantly
different from those shown in Table~1.  Fig.~\ref{chi2.fig} displays
the variation of the minimized statistical parameter with respect to
$\Delta t$ for the $g$ band light curves with each of our four
methods. In this figure microlensing and galaxy terms are not
included in the fits.

The most important conclusion to be drawn from Table~1 and
Fig.~\ref{chi2.fig} is that the best fit value of $\Delta t$ is {\em
extremely robust}, changing by less than 1\% whatever method, data
version or parameterized model is used.  The delay error estimates
produced by the various statistical analyses are somewhat more varied,
with those methods that assume greater knowledge of the statistical
character of the light curves naturally producing smaller error
estimates, but in no case is the estimated error greater than 3\%.
As we shall see in the next section, the error in the measured delay
now contributes negligibly to the error in the deduced global value of
$H_0$.

\begin{figure}
\epsscale{0.9}
\plotone{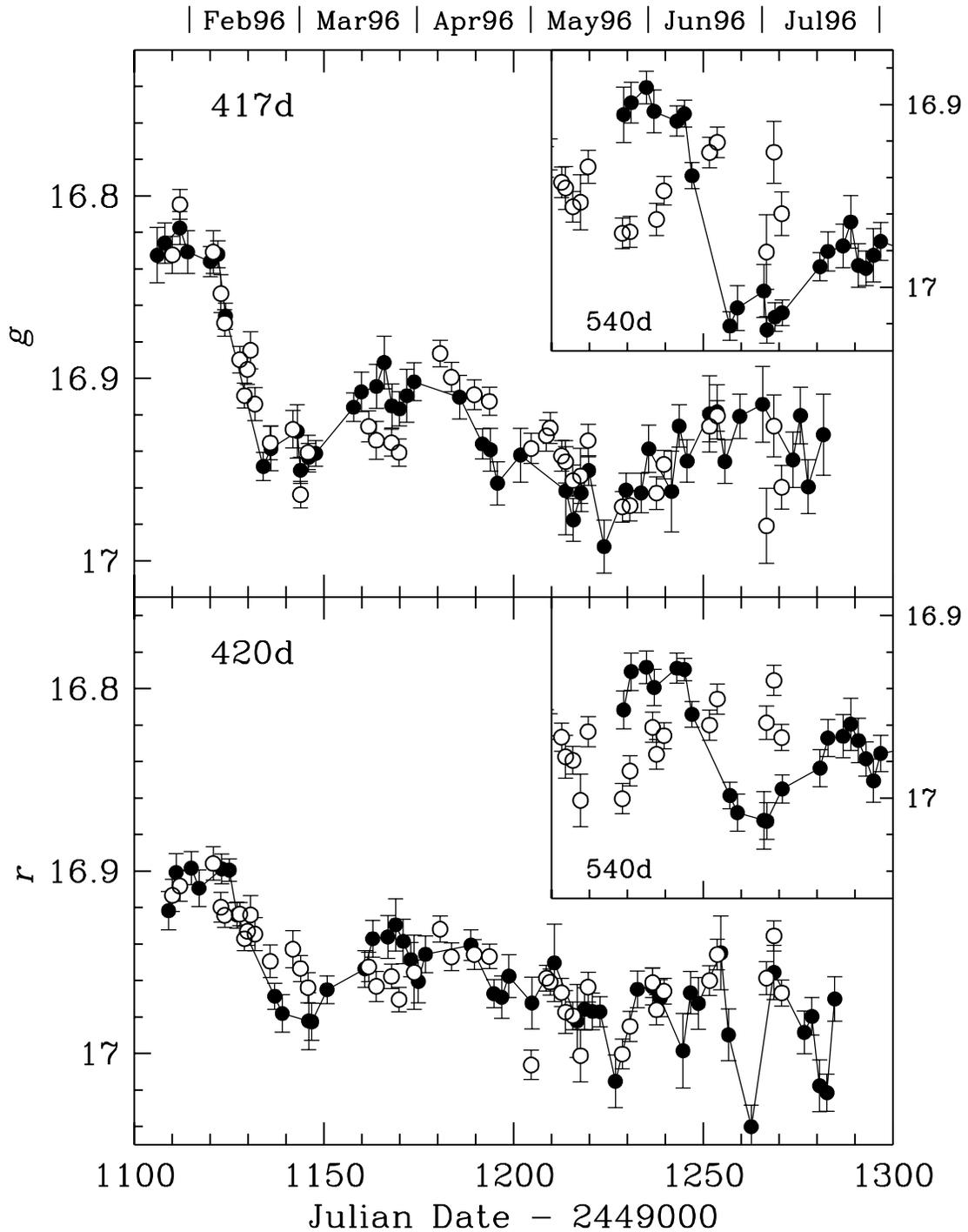}
\caption{The 1995 A light curves (filled points) shifted by the
optimal values of the time delay $\Delta t$ and the magnitude offset
$\Delta m$, superimposed on the 1996 image B data (open points).  The
fits are based on the linear method analysis, but the parameters given
by other fitting methods are nearly identical.  See the text for
details.  Insets show the overlapping regions of A and B light curves
assuming the long delay of 540 days (and fitting for the magnitude
offset). This delay is clearly excluded by the data.}
\label{linear.fig}
\end{figure}

Because it is simple and seems to work as well as any other approach
for this data set, we adopt the results of the linear method applied
to the $g$ band light curves with just a $\Delta t, \Delta m$ model as
our final numbers.  This gives a delay of $417 \pm 3$ days and a $g$
band magnitude offset of $-118^{+6}_{-8}$ millimag (B is brighter than
A).  Fig.~\ref{linear.fig} displays this fit and the corresponding
independent fit (420 days, $-215$ millimag) for the $r$ band data; it
also shows the cleanly rejected fit which corresponds to the long
delay of about 540 days in the figure insets.

\begin{figure}
\epsscale{0.9}
\plotone{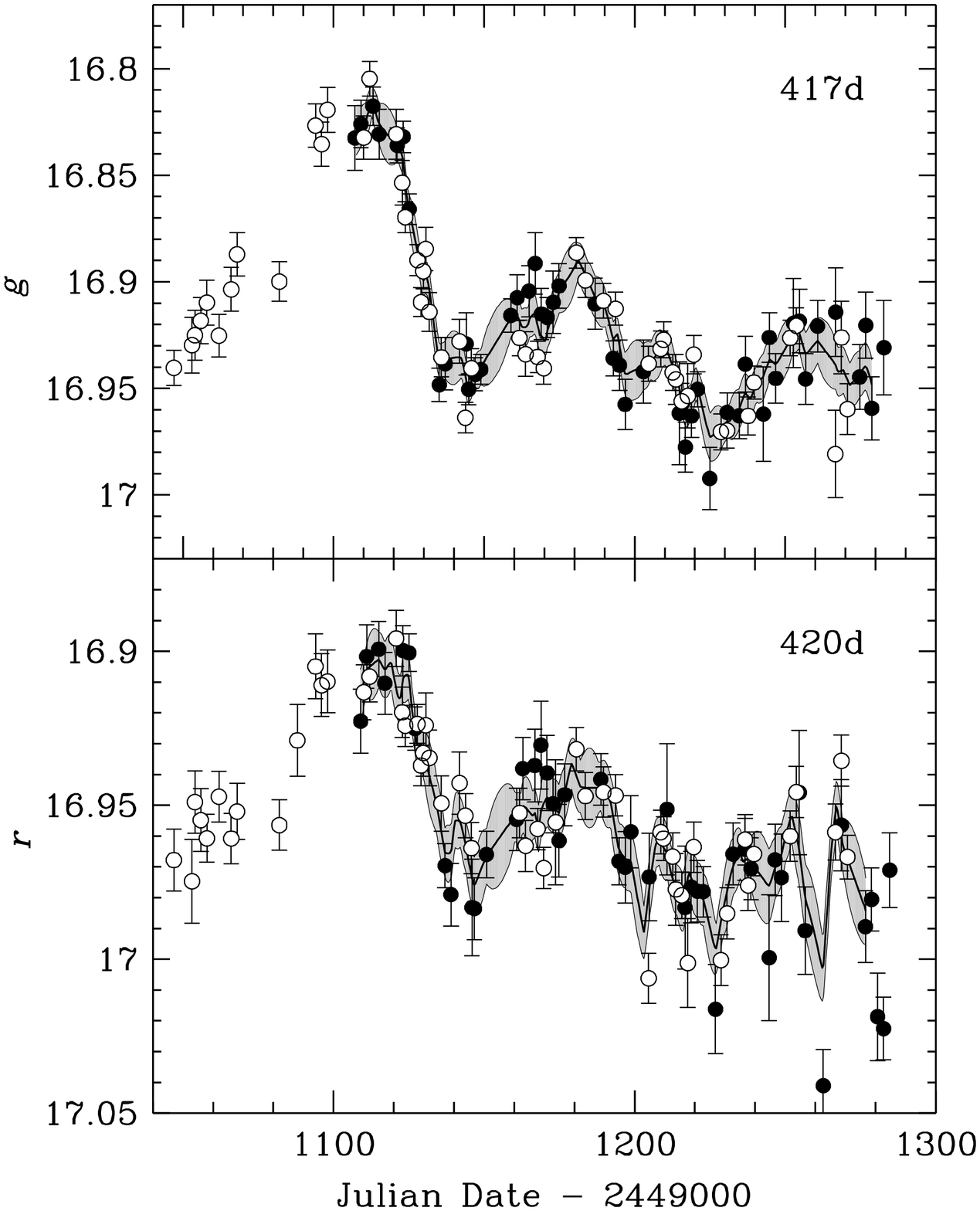}
\caption{The optimal PRH reconstruction of the shifted and combined A
(filled points) and B (open points) light curves of 0957+561. The
shaded region (``snake") corresponds to the 1$\sigma$ confidence
interval of the reconstruction.  The error bars
are the photometric 1$\sigma$ measurement errors.}
\label{PRH.fig}
\end{figure}

In addition to providing best fits and error estimates, the PRH method
produces an optimal (under the assumed statistical properties of the
source light curve) reconstruction of the underlying single light
curve sampled in the two images, and a ``snake'' corresponding to its
uncertainty at each point. This reconstruction and its comparison to
the data are shown in Fig.~\ref{PRH.fig}.

\section{A Measurement of the Global Value of $H_0$ \label{model.sec}}

For a complex gravitational lens system such as 0957+561, reliable
measurement of the differential time delay may still leave one far
from the goal of a good measurement of $H_0$ ({\em e.g.}, Schechter
\etal\ 1996), simply because of uncertainties in the lens mass
distribution and its implied conversion factor between $c \Delta t$
and the source angular diameter distance (Falco, Gorenstein \& Shapiro
1991; Kochanek 1991).  This was once a major obstacle, but fortunately
an extensive recent theoretical modeling study of 0957+561 by GN
appears to have resolved the major ambiguities with the exception of
one effect noted by Falco, Gorenstein \& Shapiro (1985); and
Gorenstein, Falco \& Shapiro (1988); see below. In particular, GN use
15 known observables to constrain a set of lens mass models having 9
free parameters.  The relation between time delay and distance is
controlled primarily by the total projected mass of the lens within
the circle defined by the diameter connecting the two source images
$M(<r)$ and by its derivative $dM/dr$ (Chang \& Refsdal 1976; Refsdal
1992; Wambsganss \& Paczy\'nski 1994).  GN find that the observed
image separation $\Delta \theta = 6.1 \arcsec$ and the observed
mapping of the complex 5-component structure in the VLBI images of the
image A and B (Garrett \etal\ 1994) nearly fix $M(<r)$ and $dM/dr$ at
the image positions, respectively, independent of other features of
the lens model.  For example, compared to their best fit model (which
models G1 as a softened power-law sphere), GN find that the best-fit
King model (after Falco \etal\ 1991) with a compact nucleus gives only
a 3\% higher value of $H_0$ despite introducing a central black hole
in G1 with a mass of $2.5 \times 10^{10} M_\odot$.  Including
ellipticity in G1 at the observed value of $\epsilon = 0.3$ and major
axis position angle of $55\arcdeg$ (Bernstein, Tyson \& Kochanek 1993)
increases $H_0$ by 2\%.  Adding perturbations from the two closest
cluster galaxies reduces it by 4\%.  Moreover, changing $\Omega_0$
from unity, the value otherwise assumed throughout this paper, to 0.1
increases $H_0$ by only 7\%.
Introducing the cosmological constant while keeping the universe flat
results in an increase of just 4\% in the $\Omega_0 = 0.25$,
$\Omega_\Lambda = 0.75$ model.

Unfortunately, the distance conversion factor supplied by the GN
models remains quite sensitive to one remaining degeneracy which
cannot be removed by the observed properties of the lensing event
itself (Falco \etal\ 1985, Gorenstein \etal\ 1988), namely how much of
the lensing is contributed by mass associated with the galaxy G1
versus how much is supplied by the associated and superimposed galaxy
cluster.  Since these two must sum to a known value (in order to
produce the observed splitting), the degeneracy may be parameterized
by either the one or the other.  GN adopt $\sigma_{\rm obs}$, the central
line-of-sight velocity dispersion of G1, to measure its contribution
and $\kappa$, the dimensionless lensing convergence contributed by the
cluster, to parameterize its effect.  In order to derive an $H_0$
value, we need to independently measure one or the other
parameter. Measuring both provides an internal consistency check.

Fischer \etal\ (1996) have recently reported a mapping of the cluster
mass distribution based on the observed distortion of faint background
galaxies, a technique which is now well developed and has been
successfully applied to several other galaxy clusters ({\em e.g.}
Tyson, Valdes \& Wenk 1990; Tyson \& Fischer 1995; Squires \etal\
1996a, b). Unfortunately, the mass profile reported by Fischer \etal\
is centered on G1, rather than the center of mass in the field located
some $22\arcsec$ northeast. Since the distortion of background
galaxies at the two quasar image positions is dominated by the mass of
G1, it is difficult to independently estimate the contribution of the
cluster. This would have been possible if the cluster profile had been
extracted around its observed center and the region around G1 excluded
from the weak lensing analysis. Assuming circular symmetry, one could
have then estimated $\kappa$ at the radius corresponding to G1 from
the uncontaminated area in the cluster.

One can still make an indirect estimate of $\kappa$ from the total
mass of the cluster, its core radius and the location of G1 with
respect to the cluster center. Following GN and Kochanek (1991), we
assume that the cluster potential corresponds to a softened isothermal
sphere $\phi(r) = b_{\rm crit} (r^2 + r_c^2)^{1/2}$, where $r_c$ is
the core radius and $b_{\rm crit} = 17\arcsec.3 (\sigma_{\rm cl}/1000
\, {\rm km \, s^{-1}})^2$ is the critical radius of the cluster
expressed in terms of its velocity dispersion $\sigma_{\rm cl}$. This
potential results in a density profile slightly different from that of
Fischer \etal\ (1996), but well within their observational
uncertainty. The local convergence at a position $r$ in the cluster is
then given by $\kappa = \frac{1}{2} b_{\rm crit} (r^2 + 2 r_c^2) /
(r^2 + r_c^2)^{3/2}$. Using direct spectroscopy of cluster galaxies,
Angonin-Willaime, Soucail \& Vanderriest (1994) find $\sigma_{\rm cl}
= 714 \pm 130$ km/s. The cluster mass estimate from weak lensing, $M(r
< 1 \, {\rm Mpc}) = 3.9 \pm 1.2 \times 10^{14} M_\odot$ (Fischer
\etal\ 1996), implies a slightly higher velocity dispersion of
$\sigma_{\rm cl} = 730 \pm 120$ km/s.  We thus adopt $\sigma_{\rm cl}
= 720 \pm 250$ km/s as our 95\% confidence region. Using $r =
22\arcsec$ for the distance of G1 from the center of the cluster and
$r_c = 5\arcsec$ for the cluster core radius (as measured by Fischer
\etal), our final estimate for the cluster surface density at the
position of the lensing event is $\kappa = 0.22 \pm 0.14$
(2$\sigma$). This result is insensitive to the precise value of $r_c$,
but it could increase considerably if G1 were much closer to the
center of the cluster. At that point, however, the quadratic
approximation of the cluster potential (which only includes the
convergence and shear) would break down as well.

Thus, including first only the quoted errors in the GN model and those of
our time delay and then adding the dominant contribution from the $\kappa$
uncertainty, we find
\begin{equation}
H_0 = 64^{+4.5}_{-5.3} \; \left( \frac{1 - \kappa}{0.78} \right) \; {\rm
km \, s^{-1} \, Mpc^{-1}} = 64^{+12}_{-13} \; {\rm
km \, s^{-1} \, Mpc^{-1}}
\end{equation}
where the errors reflect the total 95\% confidence interval.  The
primary remaining systematic uncertainty in this calculation is
associated with the mean redshift of the background galaxy population
used to map the cluster mass distribution, somewhat arbitrarily assumed
to be 1.2 by Fischer \etal\ (1996).
If it were allowed to
vary between 0.75 and 3, the derived value of $H_0$ would range
between 58 and 68 km/s/Mpc, respectively.  Considering very extreme
possibilities, $H_0$ would vary from 37 to 70 km/s/Mpc for mean
background galaxy redshifts between 0.5 and infinity.

Falco \etal\ (1997, as quoted by M. Davis 1996, private communication)
recently obtained a new measurement of the velocity dispersion in G1
from a high signal-to-noise Keck LRIS spectrum.  Although there are
some puzzling features of their data, a surprisingly rapid variation
in $\sigma_{\rm obs}$ along the slit in particular, they obtain a
value roughly in the range $\sigma_{\rm obs} = 275 \pm 30$ km/s
(2$\sigma$).  This is much more accurate than---but consistent
with---the only previously reported value of $303 \pm 50$ km/s (Rhee
1991), which was based on a significantly lower signal-to-noise
spectrum. Inserting this new value of $\sigma_{\rm obs}$ and again
propagating all the relevant errors, we find
\begin{equation}
H_0 = 64^{+7.5}_{-9.8} \; \left( \frac{\sigma_{\rm obs}}{275 \; {\rm km
\, s^{-1}}} \right)^2 \; {\rm km \, s^{-1} \, Mpc^{-1}} = 64^{+13}_{-14} \; {\rm
km \, s^{-1} \, Mpc^{-1}} \quad .
\end{equation}
where we again quote 2$\sigma$ error intervals.  The GN model error,
included in our $H_0$ uncertainty, allows for finite aperture effects
and possible anisotropy of stellar orbits.

It is important and reassuring that these two entirely independent
methods of resolving the galaxy-cluster degeneracy in 0957+561 give
very similar and entirely consistent results.  For example, if one
assumes an extremely low value of 0.5 for the mean redshift of
background galaxies in the $\kappa$ method, this would yield a very
small value of $H_0$ (see above), but would also contradict the Falco
\etal\ (1996) result by implying a G1 velocity dispersion of only 210
km/s.

We cannot properly average (2) and (3) since they are based on the
same lens mass model ({\em i.e.}, are not entirely independent), but
they can be approximately summarized by $H_0 = 64 \pm 13$ km/s/Mpc at
95\% confidence, or $\pm$ 10\% at 1$\sigma$.

\section{Summary and Discussion \label{summary.sec}}

Our main results may be summarized as follows: The sharp photometric
event predicted in the 1996 0957+561B light curve in Paper I based on
the observed 1995 image A light curve has been observed at the time
corresponding to the so called short (about 420 day) delay, thus
confirming its validity.  No such event was observed at a time
corresponding to the long (540 day) delay, thus rejecting it.  The
time delay determined by the data presented here and in Paper I is
quite robust and has a best fit value of $417 \pm 3$ days.  Combining
this value with the latest theoretical models of the lens mass
distribution plus two independent measurements of the relative
contributions of the galaxy G1 and the associated galaxy cluster
indicates that the global value of $H_0$ lies in the interval 51 to 77
km/s/Mpc with a most probable value in the range 58--70
km/s/Mpc. Although the possibility of significant systematic error
remains, especially in the lens mass distribution model, {\em this result
is likely as accurate and reliable as the best
conventional measures of} $H_0$.

Though still a significant potential source of systematic error,
there is reason to believe that further refinements of the lens
mass distribution model will have little effect on the derived value of $H_0$,
as argued in detail at the beginning of \S~\ref{model.sec}, following
GN.  The point is that the constraints provided by the detailed matching
of the A and B VLBI images (Garrett \etal\ 1994) appear to sufficiently
constrain the critical physical properties of lens models, particularly
the gradient of the projected lens mass distribution at the image positions,
that even models which differ substantially in their other properties will
yield nearly the same value of $H_0$ (Chang \& Refsdal 1976; Refsdal
1992; Wambsganss \& Paczy\'nski 1994).  These critical VLBI
constraints were not available to earlier modelers, {\it e.g.,} Falco
\etal\ (1991), who derived values of $H_0$ varying from the GN results
by up to 50\% for a fixed delay and G1 velocity dispersion.
Of course, further exploration of the space of all possible lensing models
of 0957+561A,B will be useful to determine just how much the derived
$H_0$ value can be changed within the constraints of the VLBI maps; the
extensive modeling by GN was not exhaustive of all possibilities.

The data presented here have some interesting implications beyond
those for the extragalactic distance scale.  We now turn briefly to
these.

First, note in Figures 3 and 4 that the time shifted and magnitude
offset light curves of images A and B do not appear to match
perfectly, even though they are strikingly similar.  In particular,
the image B points fall a few hundredths of a magnitude below (fainter
than) those of the shifted and offset image A data during the interval
JD 2450160 to 2450175 and then even more slightly above them during
2450175 to 2450220.  We have made no test of the statistical
significance of this discrepancy, but no acceptable delay and offset
removes it, nor have we been able to identify any plausible
shortcoming or anomaly in the data for either image during the
intervals in question which might be blamed for a spurious effect.  It
is tempting to speculatively attribute the mismatch to microlensing
perturbations of the macrolensing event since one expects a
substantial microlensing optical depth for at least image B and since
claims of such effects have been made before based on earlier light
curve data (Schild \& Smith 1991, Schild \& Thomson 1995, Schild
1996).  The timescale of the mismatch (tens of days) is far shorter
than that expected for strong microlensing perturbations by roughly
stellar mass objects, or even massive planets, in the low optical
depth limit (years to decades, Young 1981, Gott 1981), but these are
not large amplitude events and the low optical depth limit probably
does not apply.  The small magnitude of these short time scale
mismatches and the null detection of a slow offset drift $\mu$, see
Table~1, may be more conservatively interpreted as {\em upper limits}
on microlensing effects in the present data.  We may return to this
issue in a separate paper.  In any case, these minor and marginally
detected features do not appear to compromise or complicate the time
delay measurement.

Second, the shifted and offset light curves of images A and B combine
to give an intriguing record of the quasar source's intrinsic
variability, possible small microlensing perturbations aside.  See
Figures 3 and 4.  Such data may contain important clues to the
quasar's detailed internal physics, the functioning of its accretion
disk in the conventional AGN model.  Although we have not yet carried
out any such analyses, it is apparent that one might ask many
interesting statistical question of this and similar quasar light
curves.

Third, Peebles \etal\ (1991) and Dar (1991) point out that
gravitational lensing events confirm that high-redshift galaxies and
quasars are at cosmological distances.  The precise repetition of the
1995 image A light curve by image B in 1996 (to roughly 1\% accuracy)
based on an {\em a priori} prediction provides the best such available
case.  It both unambiguously proves that 0957+561A,B is a {\em bona
fide} case of gravitational lensing and gives an order-of-magnitude,
model-independent measure of the source distance $\sim (4 c \Delta t)/(\Delta
\theta^2) = 1.6$~Gpc.

Turning finally to a more general perspective, 0957+561A,B appears to
have provided a first and so far best success for Refsdal's method of
measuring $H_0$, but it is clear that there are good prospects of both
improving its accuracy and extending its applications to other
gravitational lens systems.  Towards the former goal, improved
measurements of G1's position and velocity dispersion $\sigma_{\rm
obs}$ and of the galaxy cluster's mass distribution are clearly
possible; some are already underway.  Direct measurement of redshifts
of the distorted background galaxy images will be challenging but
would remove a major systematic uncertainty.  Further theoretical
exploration of the lens mass model space is also desirable to verify
the uniqueness of the GN model.  Rapid progress can also be expected
using other lens systems.  Recently, Schechter \etal\ (1996) have
reported a convincing, though only moderately accurate, pair of time
delays in the 1115+080A,B,C system.  The accuracy can almost certainly
be improved.  It remains to be seen whether or not sufficiently unique
and well constrained models of this system can be obtained to allow
useful $H_0$ measurements.  Systems such as B0218+357, for which a
very rough estimate of the delay has been obtained from radio
polarimetry (Corbett \etal\ 1996) should be powerful for distance
determinations because their resolved structure provides strong model
constraints.  The rate of discovery of new lens systems is likely to
increase, so we may expect yet more suitable ``Rosetta Stone" systems
to be found in the future.

As noted in \S~\ref{intro.sec}, no known or imagined effect or
systematic error would cause different lens systems to give {\em the
same} wrong answer consistently.  Thus, if good time delay
measurements and well constrained lens models yield the same value of
$H_0$ for a few different lens systems (with different angular sizes,
lens and source redshift, {\em et cetera}), the problem of measuring
the global value of Hubble's constant will have been effectively
solved.

\acknowledgments

We gratefully acknowledge assistance from and informative discussions
with R. Blandford, M. Davis, D. Haarsma, J. Hewitt, R. Narayan,
B. Paczy\'nski, U. Pen, W. Press, R. Schild and K. Stanek.  The long
term, frequent sampling photometric monitoring program at APO would
not be possible without the hard work and dedication of its staff, for
which we are very thankful.  This work was supported by NSF grants
AST94-19400 and AST95-29120, and NASA grant NAG5-2759.  WNC would
like to thank the Fannie and John Hertz Foundation for its continued
support.

\clearpage

\begin{table}
\begin{center}

{\bf TABLE 1}

\vspace{1 cm}

\begin{tabular} {lccllll}  \tableline \tableline
method & band & fitted parameters &
\multicolumn{1}{c}{$\Delta t$}    &
\multicolumn{1}{c}{$\Delta m$}    &
\multicolumn{1}{c}{$\mu$}         &
\multicolumn{1}{c}{$m_{\rm gal}$} \\
       &      &                   &
\multicolumn{1}{c}{[day]}         &
\multicolumn{1}{c}{[millimag]}    &
\multicolumn{1}{c}{[$\mu$mag/day]}&
\multicolumn{1}{c}{[mag]}         \\
\tableline \tableline

Linear          & $g$    & $\Delta t$, $\Delta m$                &
$417^{+3}_{-3}$ & $-118^{+  6}_{-  8}$ & & \\

                &        & $\Delta t$, $\Delta m$, $\mu$         &
$417^{+5}_{-4}$ & $-118^{+ 31}_{- 27}$ & $-4^{+132}_{-153}$  & \\

                &        & $\Delta t$, $\Delta m$, $m_{\rm gal}$ &
$417^{+4}_{-4}$ & $-119^{+116}_{-  7}$ & & $>19.4$ \\
\tableline

Linear (event)  & $g$    & $\Delta t$, $\Delta m$                &
$417^{+3}_{-4}$ & $-123^{+ 13}_{- 12}$ & & \\
\tableline

Linear          & $r$    & $\Delta t$, $\Delta m$                &
$420^{+6}_{-9}$ & $-215^{+  6}_{-  8}$ & & \\

                &        & $\Delta t$, $\Delta m$, $\mu$         &
$420^{+8}_{-8}$ & $-200^{+ 35}_{- 27}$ & $-87^{+131}_{-188}$  & \\

                &        & $\Delta t$, $\Delta m$, $m_{\rm gal}$ &
$420^{+7}_{-9}$ & $-215^{+ 89}_{-  9}$ & & $>19.7$ \\
\tableline

Linear          & $gr$   & $\Delta t$, $\Delta m$                &
$418^{+4}_{-2}$  & & & \\
\tableline \tableline

PRH & $g$ & $\Delta t$, $\Delta m$
  & $417^{+1}_{-1}$ & $-117^{+  6}_{-  6}$ & & \\

  & & $\Delta t$, $\Delta m$, $\mu$
  & $417$           & $-117$    & $-39$ & \\

  & & $\Delta t$, $\Delta m$, $m_{\rm gal}$
  & $417$           & $-126$    & & $29.4$ \\
\tableline

PRH & $r$ & $\Delta t$, $\Delta m$
  & $420^{+2}_{-1}$ & $-215^{+  6}_{-  4}$ & &\\

  & & $\Delta t$, $\Delta m$, $\mu$
  & $420$           & $-245$     & $-127$ & \\

  & & $\Delta t$, $\Delta m$, $m_{\rm gal}$
  & $420$           & $-214$     & & $29.2$ \\
\tableline \tableline

PHKRS            & $g$  & $\Delta t$, $\Delta m$                &
$417^{+2}_{-4}$ & $-117^{+  5}_{- 10}$ & & \\

                & $r$  & $\Delta t$, $\Delta m$                &
$419^{+9}_{-9}$ & $-212^{+  7}_{- 12}$ & & \\

                & $gr$ & $\Delta t$, $\Delta m$                &
$417^{+4}_{-4}$ & & & \\
\tableline \tableline

Cross-correlation & $g$  & $\Delta t$ & $420^{+4}_{-4}$ & & & \\
                  & $r$  & $\Delta t$ & $422^{+8}_{-8}$ & & & \\
\tableline \tableline

\end{tabular}
\end{center}
\end{table}

\clearpage

\clearpage

\begin{references}

\reference{ASV94} Angonin-Willaime, M.-C., Soucail, G., \&
	Vanderriest, C. 1994, \aap, 291, 411

\reference{BTK93} Bernstein, G. M., Tyson, J. A., \& Kochanek,
	C. S. 1993, \aj, 105, 816

\reference{BO95} Beskin, G. M., \& Oknyanskij V. L. 1995, \aap, 304,
	341 

\reference{BN92} Blandford, R.D., \& Narayan, R. 1992, \araa, 30, 311


\reference{CR76} Chang, K., \& Refsdal, S. 1976, International CNRS
	Coll., 263, 369

\reference{CR79} Chang, K., \& Refsdal, S. 1979, \nat, 282, 561

\reference{COR96} Corbett, E. A., Browne, I. W. A., Wilkinson, P. N.,
	\& Patnaik, A. R. 1996, in Astrophysical Applications of
	Gravitational Lensing, ed. C. S. Kochanek, \& J. N. Hewitt
	(Dodrecht: Kluwer), 37

\reference{DAR91} Dar, A. 1991, \apj, 382, L1


\reference{FGS85} Falco, E. E., Gorenstein, M. V., \& Shapiro, I. I.
	1985, \apj, 289, L1 

\reference{FGS91} Falco, E. E., Gorenstein, M. V., \& Shapiro, I. I.
	1991, \apj, 372, 364  

\reference{FWS91} Falco, E. E., Wambsganss, J., \& Schneider, P. 1991,
	\mnras, 251, 698  

\reference{FBR96} Fischer, P., Bernstein, G., Rhee, G., \& Tyson,
	J. A. 1996, AJ, in press (astro-ph/9608117)

\reference{F84} Florentin-Nielsen, R. 1984, \aap, 138, L19 

\reference{GAR94} Garrett, M. A., Calder, R. J., Porcas, R. W., King,
	L. J., Walsh, D., \& Wilkinson, P. N. 1994, \mnras, 270, 457

\reference{GFS88} Gorenstein, M. V., Falco, E. E., \& Shapiro, I. I.
	1988, \apj, 327, 693 

\reference{G81} Gott, J. R. 1981, \apj, 243, 140

\reference{GN96} Grogin, N. A., \& Narayan, R. 1996, \apj, 464, 92 \& 473, 570 (GN)

\reference{HHL96} Haarsma, D. B., Hewitt, J. N., Leh\'ar, J., \&
	Burke, B. F. 1996, \apj, submitted (astro-ph/9607080)


\reference{K82} Keel, W. C. 1982, \apj, 255, 20 

\reference{K91} Kochanek, C. S. 1991, \apj, 382, 58

\reference{KUN95} Kundi\'c, T., Colley, W. N., Gott, J. R. III,
	Malhotra, S., Pen, U., Rhoads, J. E., Stanek, K. Z., \&
	Turner, E. L. 1995, \apj, 455, L5 (Paper I)

\reference{LHR92} Leh\'ar, J., Hewitt, J. N., Roberts, D. H., \&
	Burke,	B. F. 1992, \apj, 384, 453 

\reference{L81} Lloyd, C. 1981, \nat, 294, 727



\reference{NSC84} Narasimha, D., Subramanian, K., \& Chitre, S. M.
	1984, \mnras, 210, 79  

\reference{NAR91} Narayan, R. 1991, \apj, 378, L5

\reference{NAR91} Narayan, R. \& Bartelmann M. 1996, astro-ph/9606001 

\reference{OSC96} Oscoz, A., Serra-Ricart, M., Goicoechea, L. J.,
	Buitrago, J., \& Mediavilla, E. 1996, \apj, 470, L19


\reference{PSK91} Peebles, P. J. E., Schramm, D. N., Turner, E. L., 
	\& Kron, R. G. 1991, \nat, 352, 769

\reference{PHK94} Pelt, J., Hoff, W., Kayser, R., Refsdal, S., \&
	Schramm, T. 1994, \aap, 286, 775 (PHKRS)

\reference{PKR95} Pelt, J., Kayser, R., Refsdal, S., \& Schramm, T.
	1996, \aap, 305, 97 (PHKRS)


\reference{PRH92a} Press, W. H., Rybicki, G. B., \& Hewitt, J. N.
	1992a, \apj, 385, 404 (PRH)

\reference{PRH92b} Press, W. H., Rybicki, G. B., \& Hewitt, J. N.
	1992b, \apj, 385, 416 (PRH)

\reference{R64} Refsdal, S. 1964, \mnras, 128, 307

\reference{R66} Refsdal, S. 1966, \mnras, 132, 101

\reference{R92} Refsdal, S. 1992, in Gravitational Lenses,
	eds. R. Kayser, T. Schramm, \& L. Nieser, Hamburg: Springer
	Verlag, 61

\reference{R91} Rhee, G. 1991, \nat, 350, 211 

\reference{SCH96} Schechter, P. L. \etal\ 1996, \apj, submitted

\reference{S90} Schild, R. E. 1990, \aj, 100, 1771 

\reference{S90} Schild, R. E. 1996, \apj, 464, 125

\reference{SC86} Schild, R. E., \& Cholfin, B. 1986, \apj, 300, 209

\reference{SS91} Schild, R. E., \& Smith, R. C. 1991, \aj, 101, 813

\reference{ST95} Schild, R., \& Thomson, D. J. 1995, \aj, 109, 1970


\reference{SQU96a} Squires, G., Kaiser, N., Babul, A., Fahlman, G.,
	Woods, D., Neumann, D. M., B\"ohringer, H. 1996a, \apj, 461, 572

\reference{SQU96b} Squires, G., Kaiser, N., Fahlman, G., Babul, A.,
	Woods, D., 1996b, \apj, 469, 73

\reference{TCO92} Turner, E. L., Cen, R., \& Ostriker, J. P. 1992,
	\aj, 103, 1427

\reference{TYS90} Tyson, J. A., Valdes, F., \& Wenk, R. A. 1990, \apj,
	349, L1 

\reference{TF95} Tyson, J. A., \& Fischer, P. 1995, \apj, 446, L55

\reference{VSH89} Vanderriest, C., Schneider, J., Herpe, G.,
	Chevreton, M., Moles, M., \& Wl\'erick, G. 1989, \aap, 215, 1 

\reference{WCW79} Walsh, D., Carswell, R. F. \& Weymann, R. J. 1979,
	\nat, 279, 381 



\reference{WP94} Wambsganss, J., \& Paczy\'nski, B. 1994, \aj, 108,
	1156

\reference{Y81} Young, P. 1981, \apj, 244, 756 

\reference{YGK81} Young, P., Gunn, J. E., Kristian, J., Oke, J. B., \&
	Westphal, J. A. 1981, \apj, 244, 736  

\end{references}
\end{document}